\documentclass[preprint,review,12pt]{elsarticle}

\usepackage{setspace}
\usepackage[T1]{fontenc}
\usepackage[utf8]{inputenc}
\usepackage{longtable} 
\usepackage{booktabs}
\usepackage{amsmath, amsthm, amsfonts, amssymb, bm}
\usepackage{graphics}
\usepackage{adjustbox}
\usepackage{subcaption}
\usepackage{graphicx}
\usepackage{mathtools}
\usepackage{hyperref}
\usepackage{tikz}
\usetikzlibrary{quantikz2}

\newtheorem{Remark}{Remark}
\newtheorem{Heuristic}{Heuristic}
\newtheorem{Proposition}{Proposition}
\newtheorem{Corollary}{Corollary}

\DeclareMathOperator{\Aut}{Aut}
\DeclareMathOperator{\argmax}{argmax}
\DeclareMathOperator{\pred}{pred}

\DeclareMathOperator{\Sym}{Sym}

\journal{a Journal}

\begin{document}

\begin{frontmatter}

\title{On the Effects of Small Graph Perturbations in the MaxCut Problem by QAOA}

\author{Leonardo~Lavagna}
\ead{leonardo.lavagna@uniroma1.it}
\author{Simone~Piperno}
\ead{simone.piperno@uniroma1.it}
\author{Andrea~Ceschini}
\ead{andrea.ceschini@uniroma1.it}
\author{Massimo~Panella\corref{cor1}}
\ead{massimo.panella@uniroma1.it}

\cortext[cor1]{Corresponding author}

\address{Department of Information Engineering, Electronics and Telecommunications (DIET), University of Rome ``La Sapienza'', Via Eudossiana 18, 00184 Rome, Italy.}

\begin{abstract}\onehalfspacing
We investigate the Maximum Cut (MaxCut) problem on different graph classes with the Quantum Approximate Optimization Algorithm (QAOA) using symmetries. In particular, heuristics on the relationship between graph symmetries and the approximation ratio achieved by a QAOA simulation are considered. To do so, we first solve the MaxCut problem on well-known graphs, then we consider a simple and controllable perturbation of the graph and find again the approximate MaxCut with the QAOA. Through an analysis of the spectrum of the graphs and their perturbations, as well as a careful study of the associated automorphism groups, we aim to extract valuable insights into how symmetry impacts the performance of QAOA. These insights can then be leveraged to heuristically reduce the quantum circuit complexity, the number of training steps, or the number of parameters involved, thus enhancing the efficiency and effectiveness of QAOA-based solutions.
\end{abstract}


\end{frontmatter}

%
%
\section{Introduction}
\label{sect:intro}
The pursuit of quantum computing has spanned decades, with quantum algorithms offering exponential speedup for various computational tasks \cite{Mont15}. Despite challenges in noise and qubit limitations, Noisy Intermediate-Scale Quantum (NISQ) computers have emerged as state-of-the-art devices to carry out quantum computations \cite{Pre18}, generating excitement for achieving quantum supremacy \cite{Aru19}. On these bases much effort has been dedicated to developing practical algorithms, the so-called Variational Quantum Algorithms (VQAs) \cite{Ben19,Per13}, with an hybrid approach, where the parameters of a quantum circuit are updated using classical optimization routines, showing important results in quantum chemistry, machine learning, combinatorial optimization, and more \cite{Cer21}.

The Quantum Approximate Optimization Algorithm (QAOA) \cite{Far14} stands out as a promising VQA, capitalizing on the parallelism of quantum systems. QAOA is designed to efficiently find approximate solutions of combinatorial optimization tasks, usually Quadratic Unconstrained Binary Optimization (QUBO) problems \cite{Koc14},\cite{Glo19}, contributing to real-world applications, from scheduling \cite{Cho20}, \cite{Kur23}, routing \cite{Aza23} and network analysis \cite{Jin23}, to protein folding \cite{Rob21}, drug discovery \cite{Mus22} and portfolio optimization \cite{Bra22}. This kind of problems are usually defined by an objective function $z \to z^T\mathbf{q}z\in \mathbb{R}$, where $z\in \{0,1\}^n$ and $\mathbf{q}=(q_{ij})\in\mathbb{R}^{n^2}$ is a problem-dependent upper-triangular matrix for which we seek $z^*:=\argmax\{z^T\mathbf{q}z\}$.

In this paper, we focus on the paradigmatic Maximum Cut (MaxCut) problem on a graph $\Gamma=(V,E)$, where we want to find a partition $\{S,V\setminus S\}$ of the set of nodes $V$ with the largest subset $E_{cut}\subset E$ of edges connecting the nodes in $S$ to those in $V\setminus S$. In this case the objective function is given by 
\begin{equation} 
g_\Gamma(z) := \sum_{(i,j)\in E}(1-z_iz_j)\,,
\label{eq:1}
\end{equation}
where $z_i=1$ means that the $i$-th node $v_i$ is in $S$ and $z_i=0$ means that $v_i \in V\setminus S$, and where the sum is extended over all edges  connecting $v_i$ to $v_j$ for $1\leq i<j\leq n$. The optimal solution of the MaxCut problem with respect to $\Gamma$ will be denoted as $g_\Gamma(z^*)$. It is well-known that finding $g_\Gamma(z^*)$ is in general NP-hard \cite{Kor12}, nonetheless many classical approximate solutions have been found \cite{Viz03}, notably within semi-definite programming \cite{Goe95}, but such classical approaches have strong limitations that quantum algorithms are set to overcome \cite{Kar99}, \cite{Cro19}. 

Mapping every bit $z_i\in \{0,1\}$ to a Pauli-$z$ gate $\bm{\sigma}^{(i)}_z\in SU(2)$ and mapping $1$ to the identity matrix $\mathbf{1}_4\in SU(4)$ allows us to associate to the objective function an observable, denoted with a little abuse of notation as $g_\Gamma=g_\Gamma(\bm{\sigma})$, with the bit string $z$ replaced by a set of spin matrices $\bm{\sigma}$ composed of Pauli-$z$ gates with eigenvalues $\pm 1$ (that is why we add a term $\frac{1}{2}$ in order to have unit contribution for each edge), of the form
\begin{equation} 
    g_\Gamma(\bm{\sigma}) := \frac{1}{2}\sum_{(i,j)\in E}(\mathbf{1}_4-\bm{\sigma}^{(i)}_z\otimes \bm{\sigma}^{(j)}_z)\,,
    \label{eq:2}
\end{equation}
which we can process with a NISQ device following the steps of the QAOA, here summarized:
\begin{enumerate}
    \item Fix an integer $p\geq 1$ corresponding to the number of layers in the algorithm and choose basis encoding $z\to \ket{z}$. 
    \item Prepare, using Hadamard gates $H$, the uniform superposition state $\ket{\psi_0}:=H^{\otimes n}\ket{0}^{\otimes n}=\frac{1}{\sqrt{2^n}}\sum_{z}\ket{z}$,
    where the sum is extended over all standard basis elements of $(\mathbb{C}^2)^{\otimes n}$ encoding the strings $z\in \{0,1\}^n$. 
    \item For each $k=1,\dots, p$ define a cost operator $U_{g_\Gamma}(\gamma_k)=e^{-i\gamma_k g_\Gamma(\bm{\sigma})}$ and a mixer operator $U(\beta_k)=e^{-i\beta_k\sum_{j=1}^n \bm{\sigma}_x^{(j)}}$ in analogy with the transverse-field Ising model \cite{Das05} with Pauli-$x$ gates $\bm{\sigma}_x^{(j)}\in SU(2)$, where ${\gamma_k\in [0,2\pi)}$ and ${\beta_k\in [0,\pi)}$ are randomly chosen angles.
    \item Let 
    \begin{equation}
    \ket{\gamma\beta}:=U(\beta_p)U_f(\gamma_p)\prod_{i=1}^{p-1}[U(\beta_{p-i})U_{g_\Gamma}(\gamma_{p-i})]\ket{\psi_0}
    \end{equation}
    be the evolved state and consider the expected value $F_{p,\Gamma}(\gamma,\beta):=\langle \gamma\beta| g_\Gamma(\bm{\sigma})| \gamma\beta\rangle$, use a classical device to solve for $F_{p,\Gamma}(\gamma^*,\beta^*)=\max_{\gamma,\beta}F_{p,\Gamma}(\gamma,\beta)$, and get an updated set of parameters $\gamma^*=(\gamma_1^*,\dots, \gamma_p^*)$, $\beta^*=(\beta_1^*,\dots, \beta_p^*)$. 
    \item If needed, repeat the process with $p\to p'=p+1$ and a new set of random angles $(\beta,\gamma)\to(\beta',\gamma')\in [0,\pi)^{p+1}\times [0,2\pi)^{p+1}$ and stop when the approximation ratio 
    \begin{equation}
        AR(p,\Gamma):=\frac{F_{p,\Gamma}(\gamma^*,\beta^*)}{g_\Gamma(z^*)}\in [0,1]
        \label{eq:3}
    \end{equation}
    is as close as 1 as desired.
\end{enumerate}

As $p\to +\infty$ we are guaranteed that the maximum of $F_{p,\Gamma}$ converges to the optimal value $g_\Gamma(z^*)$ \cite{Far14}, but in practice the technology is not yet mature enough to let us explore this asymptotic behavior. 
Moreover, despite many positive theoretical results about QAOA circuits, such as \cite{Far14,Far15,Wan18,Wur21}, they face challenges in obtaining nontrivial performance guarantees, especially for $p\gg 1$, due to the complex underlying physics and mathematics of the problem. Rigorous bounds are mostly established for small $p$ values (e.g., $p = 1$ or $p = 2$) often in worst-case scenarios, but lack consideration of problem instance structures \cite{Gue19}, \cite{Sha19}, with some notable exceptions about the well known Grover-type quadratic speedups \cite{Jia17}, as well as some robust heuristics for certain graph classes \cite{Zho20}. General formulas seem elusive precisely beyond small $p$ values, although initial numerical studies suggest that large $p$ values are necessary to obtain a potential quantum advantage \cite{Has19}. 
Nonetheless some important heuristics have been found, more efficient QAOA variants have been explored, and fruitful connections with other VQAs have been established, shedding some light on such challenges, at least, from an experimental point of view \cite{Ble23}.  

Considering these issues in general settings, an emerging approach, inspired by analogies with problems in condensed-matter \cite{Chi16}, statistical mechanics \cite{Coj22} and related fields \cite{Ost11}, involves applying symmetry ideas to quantum computations \cite{Ala07}, in particular to QAOA applications \cite{Sha20}. Leveraging insights from symmetries shared by the QAOA quantum circuit and initial state, as well as those of the target classical optimization problem, offers potential advancements \cite{Bra20}, \cite{Her21}. Likewise, to accelerate the QAOA evaluation, the use of symmetries has been successfully exploited by associating the classical symmetries in the objective function to the symmetries in the cost operator of the QAOA, leading to provable enhanced evaluation performances and results \cite{Shay21}. Moreover, symmetries of the local Hamiltonian describing the problem in a QAOA circuit have been related to the reduction of its parameters' space dimension as well as to subtle properties of the problem such as its associated Lie algebra \cite{Cer23}. 

In this paper, we will address the challenges faced by the QAOA exploiting symmetry, with a particular focus on the MaxCut problem, where the instances are graphs and their symmetries are certain groups of automorphisms for which there is a vast literature \cite{God01} and, thanks to a recent breakthrough in the graph isomorphism problem \cite{Bab16}, there are also useful graph-automorphisms finder routines. In particular, we will study typical MaxCut instances on different graph type classes by evaluating their symmetries and spectrum and solving the problem with the QAOA on a simulator. We will then perturb the instances in a simple and controllable way, and solve again the MaxCut problem with the QAOA in order to gain insights about the relationship between problem instance characteristics (i.e. automorphism group and spectrum) and quality of the approximate solution in terms of approximation ratio. 
Specifically, we will perturb the graphs by considering the following most elementary transformations:
\begin{itemize}
\item Adding one or two shadow nodes, that is changing $\Gamma=(V,E)$ to ${\Gamma'=(V',E)}$, where ${V'=V\cup\{\overline{v}\}}$ has an extra disconnected node, or $V'=V\cup\{\overline{v},\overline{w}\}$ has two extra disconnected nodes.
\item Selectively removing one edge at a time, without discarding the associated nodes, that is changing $\Gamma=(V,E)$ to $\Gamma'=(V,E')$ where $E'=E\setminus\{e\}$ has been reduced eliminating an edge $e\in E$.
\item Adding a pendant edge, that is changing $\Gamma=(V,E)$ to $\Gamma'=(V',E')$ where $V'=V\cup \{\overline{v}\}$ and $E'=E\cup \{\overline{e}\}$ with $\overline{e}$ connecting the degree-one node $\overline{v}$ to a random node $v\in V$.
\end{itemize}
Clearly these perturbations can: change the symmetry group $\Aut(\Gamma)$ of $\Gamma$, change the spectrum $\sigma(\Gamma)$ of $\Gamma$, change the QAOA circuit corresponding to $\Gamma$ as well as changing the MaxCut value in $\Gamma$. We will evaluate the corresponding perturbed symmetry group and spectrum, and observe the induced modification in the associated QAOA circuit and the effect of this perturbations on the simulation results. With these small perturbations we will see that it is possible to either heuristically reduce the QAOA circuit complexity, or to (heuristically) enhance the classical search for the optimal QAOA parameters, and, even, find equivalent instances with low dimension whose parameters are nearly optimal for a related highly dimensional problem, generalizing some aspects of the QAOA parameter transferability strategy proposed in \cite{Gal23}.

%
%
\section{Small Perturbations of Selected Graphs and QAOA Circuit}
\label{sect:QAOA_circuit}
\begin{figure}[!ht]
\centering
\includegraphics[width=0.8\columnwidth]{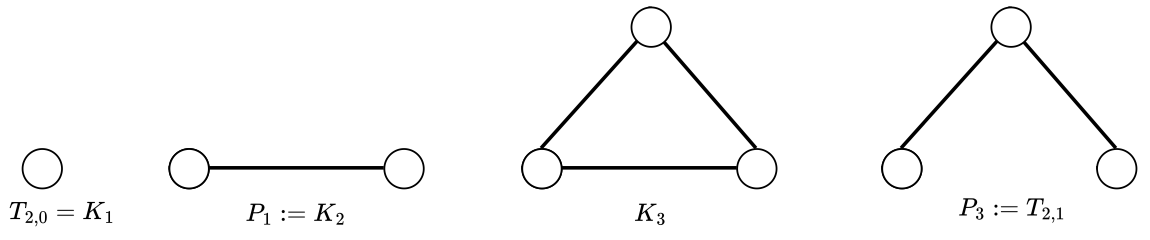}
\caption{Conventional graphs $K_1,K_2,K_3$, and $T_{2,1}$.}
\label{Fig. 2}
\end{figure}

In the spirit of using symmetry ideas to gain insights about the relationship between QAOA performances in terms of approximation ratio and graph characteristics in the MaxCut problem, we will focus on small perturbations of the problem instance, as discussed in the previous section. Here we analyze the effect of shadow nodes, pendent edges, and deleted edges on the spectrum and automorphism group of the graphs we will use in the simulations, which are representative of the main problems addressed in the literature and in the applications. These instances belong to:
\begin{enumerate}
    \item The class $K_n$ of complete graphs with $n$ nodes, that have connections among every pair of nodes.
    \item The class $E_{q,n}$ of Erd\H{o}s-R\'enyi graphs with an edge probability $q=0.5$, that have random edges assigned with probability $q$ to node pairs.
    \item The class $T_{2,h}$ of rooted full binary trees with height $h$ and $n+1=2^h$ nodes,  in which every parent node or internal node has either two or no children.
    \item The class $R_{d,n}$ of (random) regular graphs with degree $d=3$ and $n$ nodes, where the nodes have all the same degree $d$ (and one such graph is selected at random among the family of $d$-regular graphs).
\end{enumerate}
We will discuss with more details these instances in the next section, here let us set some conventions: $K_0=\emptyset$ is the null graph (with zero nodes and no edges), $K_1$ is the empty graph with one node and no edges, $P_1:=K_2$ is a simple path of length one given by a single edge, $T_{2,0}=K_1$, and $P_3:=T_{2,1}$ is a simple path of length two given by two consecutive edges as in Fig.~\ref{Fig. 2}.
\begin{Remark}\label{Rem. 1}\normalfont 
If not stated otherwise all the graphs considered will be finite, simple, connected and undirected.
\end{Remark}

%
\subsection{Spectrum of the selected graphs and their perturbations}
Recall that if $\mathbf{a}(\Gamma)$ is the adjacency matrix of a graph $\Gamma=(V,E)$, its spectral decomposition is $\mathbf{a}(\Gamma)=\sum_{i=1}^N \lambda_i \mathbf{p}_i$,
where $\lambda_i$ is an eigenvalue of $\mathbf{a}(\Gamma)$ with corresponding eigenspaces $ES(\lambda_i)$, where $\mathbf{p}_i$ is the orthogonal projector onto $ES(\lambda_i)$, and where $N$ is the number of distinct eigenvalues of $\mathbf{a}(\Gamma)$. The collection of the eigenvectors of $\mathbf{a}(\Gamma)$ is the spectrum $\sigma(\Gamma)$ which is completely determined by the characteristic polynomial $\phi_\Gamma(\lambda):=\det(\mathbf{a}(\Gamma)-\lambda\mathbf{1}_n)$, where $\mathbf{1}_n$ is the $n\times n$ identity matrix, with $n=|V|$. 
\begin{Proposition}\label{Prop. 1} The addition of $s>0$ shadow nodes to a graph $\Gamma$ generates a graph $\Gamma'$ with $n+s$ nodes containing $\Gamma$ whose characteristic polynomial is $\phi_{\Gamma'}(\lambda)=(-\lambda)^s\phi_{\Gamma}(\lambda)$.
\end{Proposition}
\noindent\textit{Proof.} For $s=1$, it holds
\begin{equation}
	\mathbf{a}(\Gamma')=\begin{pmatrix} x & y^T \\ \epsilon & \mathbf{a}(\Gamma)\end{pmatrix}\,,
\end{equation}
where $\epsilon\in\{0,1\}$, $x=(x_1,\dots, x_{n})^T\in \{0,1\}^{n}$, and ${y^T=(y_1,\dots, y_{n})\in \{0,1\}^{n-1}}$. 
Since \cite{Sil00}:
\begin{equation}
	\det(\mathbf{a}(\Gamma'))=x-y^T\mathbf{a}(\Gamma)^{-1}\epsilon\det(\mathbf{a}(\Gamma))\,,
\end{equation}
being a shadow node disconnected from all the other nodes, we have $x_i=0$ for each $i=1,\dots, n$, $y_j=0$ for each $j=1,\dots, n-1$, and $\epsilon=0$. In particular, provided that the graph is non empty,
it holds
\begin{equation}
\det(\mathbf{a}(\Gamma')-\lambda\mathbf{1}_{n+1}) =\Bigl((x-\lambda)-y^T(\mathbf{a}(\Gamma)-\lambda\mathbf{1}_n)^{-1}\epsilon\Bigr)\det(\mathbf{a}(\Gamma)-\lambda\mathbf{1}_n)\,,
\end{equation}
so $\phi_{\Gamma'}(\lambda)=-\lambda \phi_\Gamma(\lambda)$, and the thesis follows by induction on the number of shadow nodes. \hfill $\Box$
\begin{Proposition}\label{Prop. 2}\cite{Row96} Removing an edge $(u,v)$ from a graph $\Gamma$ generates a graph $\Gamma-uv$ contained in $\Gamma$ whose characteristic polynomial is 
\begin{equation}
\phi_{\Gamma-uv}(\lambda)=\phi_\Gamma(\lambda)-\phi_{\Gamma-u-v}(\lambda)+2\phi_{\Gamma}(\lambda)\sum_{i=1}^t\frac{p^{(i)}_{uv}}{\lambda-\lambda_i}\,,
\end{equation}
where $\Gamma-u-v$ is the graph obtained from $\Gamma$ by deleting node $u$, node $v$, and all the edges connected to them, and where $p^{(i)}_{uv}$ is the $(u,v)$-entry of the projector $\mathbf{p}_i$ in the spectral decomposition of $\mathbf{a}(\Gamma)$. 
\end{Proposition}

Note that the calculations needed to get $\phi_{\Gamma-u-v}(\lambda)$ are not as straightforward as in Proposition \ref{Prop. 1} since the deletion of a set nodes affects also the associated incident edges. We just mention a well known formula to compute $\phi_{\Gamma-u-v}(\lambda)$. 
\begin{Proposition}\label{Prop. 3}\cite{Row96} With the notations as in Proposition \ref{Prop. 2}, it holds:
\begin{equation}
\phi_{\Gamma-u-v}(\lambda)=\phi_\Gamma(\lambda) \Biggl[\sum_{i=1}^t\frac{p^{(i)}_{uu}}{\lambda-\lambda_i}\sum_{i=1}^t\frac{p^{(i)}_{vv}}{\lambda-\lambda_i}-\Biggl(\sum_{i=1}^t\frac{p^{(i)}_{uv}}{\lambda-\lambda_i}\Biggr)^2\Biggr]\ .
\end{equation}
\end{Proposition}

We have discussed the effect of adding shadow nodes and the effect of deleting an edge, now we will see the effect of extending a graph with a pendent edge  (i.e. an edge where there is at least a node of degree one) which in turn will also tell us something about specific formulas for the spectrum of some graphs of interest.
\begin{Proposition}\label{Prop. 4}\cite{Har71} If $\Gamma_u$ is the graph obtained from $\Gamma$ by adding a pendant edge at vertex $u$, then $\phi_{\Gamma_u}(\lambda)=\lambda\phi_{\Gamma}(\lambda)-\phi_{\Gamma-u}(\lambda)$, where $\Gamma-u$ is the graph obtained from $\Gamma$ by removing node $u$ and all the edges connected to it.
\end{Proposition}
From the previous result it is possible to obtain the spectrum of a tree, not necessarily binary. Consider a rooted tree $\Gamma=(V,E)$ with root $\omega\in V$, where the successor of every node $v\in V$ is denoted as $\succ(v)$, and the predecessor as $\pred(v)$. Define for any node $u\ne \omega$ the subgraph $C(u)$ as the component of $\Gamma-uv$ containing $u$ when $v=\pred(u)$, define also 
\begin{equation}
C'(u)=C(\succ(u)_1)\cup C(\succ(u)_2)\cup \dots \cup C(\succ(u)_k)\,,
\end{equation}
where $k$ is the number of successors of $u$. Clearly $C(\omega)=\Gamma$ and if $u$ is a leaf $\phi_{C(u)}(\lambda)=\lambda$ and $\phi_{C'(u)}(\lambda)=1$. By the well known fact \cite{Cve09} that the spectrum factorizes over the components of a graph, we have for any forest $F$, the formula 
\begin{equation}
\phi_F(\lambda)=\lambda\phi_{F-u}(\lambda)-\sum_{i=1}^d\phi_{F-u-v_i}(\lambda)\,,
\end{equation}
where the $v_i$s are the $d$ neighbors of a node $u$ in $F$. Applying this formula to the forest $C(u)$ yields the following result.
\begin{Corollary}\label{Cor. 1}
With the above notations:
\begin{equation}
\phi_{C'(u)}(\lambda)=\prod_{i=1}^d\phi_{C(\succ(u)_i)}(\lambda)
\end{equation}
and 
\begin{equation}
\phi_{C(u)}(\lambda)=\phi_{C'(u)}(\lambda)\Bigl (\lambda-\sum_{i=1}^d\frac{\phi_{C'(\succ(u)_i)}(\lambda)}{\phi_{C(\succ(u)_i)}(\lambda)}\Bigr )\,.
\end{equation}
\end{Corollary}

By a recursive application of the formulas in Corollary \ref{Cor. 1} we can get $\phi_{T_{2,h}}(\lambda)$ starting from the nodes at level $h$ and going up, determining at each step $\phi_{C'(\succ(u)_i)}(\lambda)$ and $\phi_{C(\succ(u)_i)}(\lambda)$, with a linear-time algorithm \cite{Moh89}.
\begin{Proposition}\label{Prop. 5} \cite{Cve09} If $\Gamma=(V,E)$ is a $r$-regular graph with $n$ nodes, and $\overline{\Gamma}=(V,\overline{E})$ is its complement (where $(u,v)\in \overline{E}$ if and only if $(u,v)\not \in E$), then 
$\phi_{\overline{\Gamma}}(\lambda)=(-1)^n(\lambda-n+r+1)(\lambda+r+1)^{-1}\phi_{\Gamma}(-\lambda-1)$.
\end{Proposition}
%
%
\begin{Corollary}\label{Cor. 2} For $n>1$ it holds: 
\begin{equation}
\phi_{K_n}(\lambda)=(\lambda-n+1)(\lambda+1)^{n-1}\,.
\end{equation}
\end{Corollary}
\noindent\textit{Proof.} The empty graph with $n$ nodes has characteristic polynomial $(-\lambda)^n$, being $K_n$ its complement, the thesis follows from Proposition \ref{Prop. 5}. \hfill $\Box$
\\

We have characterized $\sigma(\Gamma')$ for some perturbation $\Gamma'$ in terms of $\sigma(\Gamma)$, where we derived explicit formulas when $\Gamma=K_n$ and $\Gamma=T_{2,h}$. In the other two cases of interest $\Gamma=R_{3,n}$ and $\Gamma=E_{q,n}$ we cannot do the same due to the randomness involved in their construction. Ad hoc computations will be discussed in the next section.

\subsection{Automorphisms of the selected graphs and their perturbations}
\begin{figure}[!ht]
\begin{center}
\includegraphics[width=0.65\columnwidth]{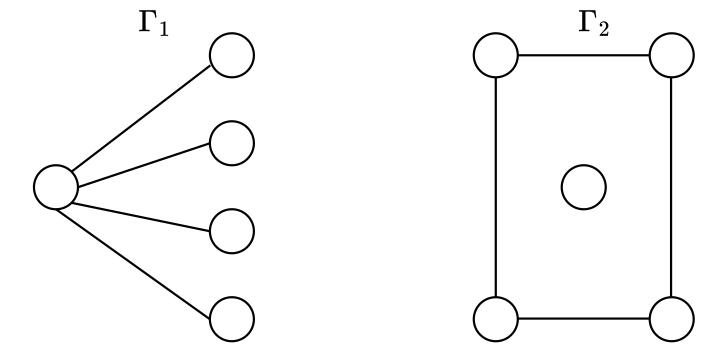}
\caption{Two non-isomorphic co-spectral graphs.}
\label{Fig. 3}
\end{center}
\end{figure}

Recall that an isomorphism between two graphs is an edge-preserving bijective map among them that we call automorphism when the two graphs coincide. An automorphism $\alpha$ of a graph $\Gamma=(V,E)$ can be seen as permutation of its nodes such that $(u,v)\in E$ if and only if $(\alpha(u),\alpha(v))\in E$. The group $\Aut(\Gamma)<\Sym(V)\cong S_{|V|}$ is a subgroup of the symmetric group acting on $V$ (where $\Sym(V)$ is isomorphic to the finite group $S_{|V|}$ of permutations acting on $\{1,\dots, |V|\}$). Recall, also, that  $\alpha\in \Aut(\Gamma)$ if and only if the adjacency matrix $\mathbf{a}(\Gamma)$ of $\Gamma$ satisfies $\mathbf{a}(\Gamma)=\bm{\delta}_\alpha^T\mathbf{a}(\Gamma)\bm{\delta}_\alpha$, where, for each $i,j\in\{1,\dots, n\}$, the matrix $\bm{\delta}_\alpha=(\delta_{\alpha(i)j})$ is the permutation matrix associated to $\alpha$ with entries $\delta_{\alpha(i)j}=1$ if the $\alpha(i)$-th node is different from the $j$-th node, and $\delta_{\alpha(i)j}=0$ otherwise.
%
\begin{Proposition}\label{Prop. 6} For any graph-isomorphism $f$ it holds $\sigma(\Gamma)=\sigma(f(\Gamma))$, and we say that the two graphs are cospectral. Viceversa, there exists non-isomorphic cospectral graphs. 
\end{Proposition}
\noindent\textit{Proof.} This is a well known fact \cite{Cve09} that we show for completeness. One implication is trivial, viceversa the two graphs $\Gamma_1$ and $\Gamma_2$ in Fig.~\ref{Fig. 3} share the same spectrum $\{0,\pm 2\}$, but there is no edge-preserving bijective map $\Gamma_1\to \Gamma_2$ since $\Gamma_1$ is connected, while $\Gamma_2$ is not. \hfill $\Box$
\begin{Proposition}\label{Prop. 7} If $\Gamma=(V,E)$, and $\Gamma'=(V',E)$ is obtained from $\Gamma$ adding $s\geq 1$ shadow nodes, then $\Aut(\Gamma')\cong S_s\times \Aut(\Gamma)$, with $S_1\times \Aut(\Gamma)\cong \Aut(\Gamma)$.
\end{Proposition}
\noindent\textit{Proof.} If $s=1$ and $\overline{v}$ is the added shadow node, since by definition $\overline{v}$ is disconnected from every other node $v\in V$, every $\alpha'\in \Aut(\Gamma')$ acts on $V'=V\cup\{\overline{v}\}$ by permutation of the elements $v\in V$ and fixes $\overline{v}$. In particular the restriction $\alpha'_{|V}$ of $\alpha'\in \Aut(\Gamma')$ to $V$ yields an automorphism of $\Gamma$, and $\Aut(\Gamma)\cong \Aut(\Gamma')$. When $s=2$ we can reason analogously. If $V'=V\cup\{\overline{v},\overline{w}\}$ is the set of nodes of $\Gamma'$, then $\alpha'\in \Aut(\Gamma')$ will permute every $v\in V$ and fix $\{\overline{v},\overline{w}\}$, where either $\alpha'(\overline{v})=\overline{w}$ and $\alpha'(\overline{w})=\overline{v}$, or $\alpha'(\overline{v})=\overline{v}$ and $\alpha'(\overline{w})=\overline{w}$. In the general case in $\Aut(\Gamma')$ there is a transposition $\tau$ interchanging $\overline{v}$ and $\overline{w}$ which yields a factor isomorphic to $S_2$, and all the the remaining permutations yield a factor isomorphic to $\Aut(\Gamma)$. The general case $s>2$ is obtained with the same arguments, replacing $\tau\in S_2$ with $\tau'\in S_s$. \hfill $\Box$ 
\begin{Remark}\label{Rem. 2}\normalfont 
If we perturb a graph by adding a pendant edge or by removing an edge, we cannot say much about the effect of the perturbation on its symmetry group. For example in Figure \ref{Fig. 3} we have a star graph $\Gamma_1$ whose symmetry group is isomorphic to $S_4$ since any automorphism will fix the degree-four node and permute the other four degree-one nodes, but removing an edge yields a star graph with a node of degree three, then three nodes of degree one and a disconnected node, so the resulting automorphism group is isomorphic to $S_3$. Similarly, adding a pendant edge to the highest-degree node in $\Gamma_1$ yields a star graph with $5$ leaves of degree one and a node of degree 5, whose symmetry group is isomorphic to $S_5$. On the other hand, in Figure \ref{Fig. 3}, the graph $\Gamma_2$ is a square plus a shadow node, so its symmetry group is isomorphic to the dihedral group $D_4\cong S_4 \rtimes S_2$ obtained as a semidirect product. If we remove an edge from $\Gamma_2$ we obtain a path of length three plus one shadow node with symmetry group isomorphic to $S_2$ (there is only one non-trivial permutation acting on the two degree-one nodes). If we add a pendant edge on the shadow node in $\Gamma_2$, instead, we get a symmetry group isomorphic to $S_2\times D_4$ since the two connected components of such perturbed graph must remain fixed by an automorphism, generating a factor isomorphic to $D_4$, and the two shadow nodes can be interchanged by a factor isomorphic to $S_2$ as in Proposition \ref{Prop. 7}.
\end{Remark}
Despite the difficulties in finding the automorphisms of a perturbed graph with an extra pendant edge or with a deleted edge, for some graphs of interest we can say something more.
\begin{Proposition}\label{Prop. 8} The automorphism group of $T_{2,h}$ is given by iterated wreath products of $S_2$. In particular $\Aut(T_{2,2})=(S_2\times S_2)\rtimes S_2=S_2\wr S_2$ and $\Aut(T_{2,3})=(\Aut(T_{2,2})\times \Aut(T_{2,2}))\rtimes S_2=(S_2\wr S_2)\wr S_2$.
\end{Proposition}
\noindent\textit{Proof.} This is a well known fact 
that we show for completeness. Let $G_h:=\Aut(T_{2,h})$, clearly $G_1\cong S_2$. Also $G_2\cong (S_2\times S_2)\rtimes S_2 $ since there are automorphisms $\alpha_1,\alpha_2$ interchanging the pair of leaves on the left and the pair of leaves on the right, generating $\langle \alpha_1,\alpha_2\rangle\cong S_2\times S_2$, and there is the additional automorphism $\alpha_3$ interchanging the branches, so $G_2\cong \langle \alpha_1,\alpha_2,\alpha_3\rangle $ with $S_2\times S_2 \lhd G_2$ a normal subgroup. For $h>2$, by the same reasoning, we have the recursive formula $G_{h+1}\cong G_h\wr S_2$.  \hfill $\Box$
%
\begin{Proposition}\label{Prop. 9} Deleting one edge from $T_{2,h}$ at level $0<r\leq h$ generates a graph $\Gamma'$ contained in $T_{2,h}$ such that:
\begin{enumerate}
    \item If $r=h$ then $\Aut(\Gamma')\cong \Aut(T_{2,1})\times \Aut(T_{2,2})\times \dots \times \Aut(T_{2,h-1})$.
    \item If $r\leq h-1$ then $\Aut(\Gamma')\cong \Aut(T_{2,r})\times \Aut(T_{2,r})\times \Aut(T_{2,r+1})\times \dots \times \Aut(T_{2,h-1})$.
\end{enumerate}
\end{Proposition}

\begin{figure}[!ht]
\centering
\includegraphics[width=0.75\columnwidth]{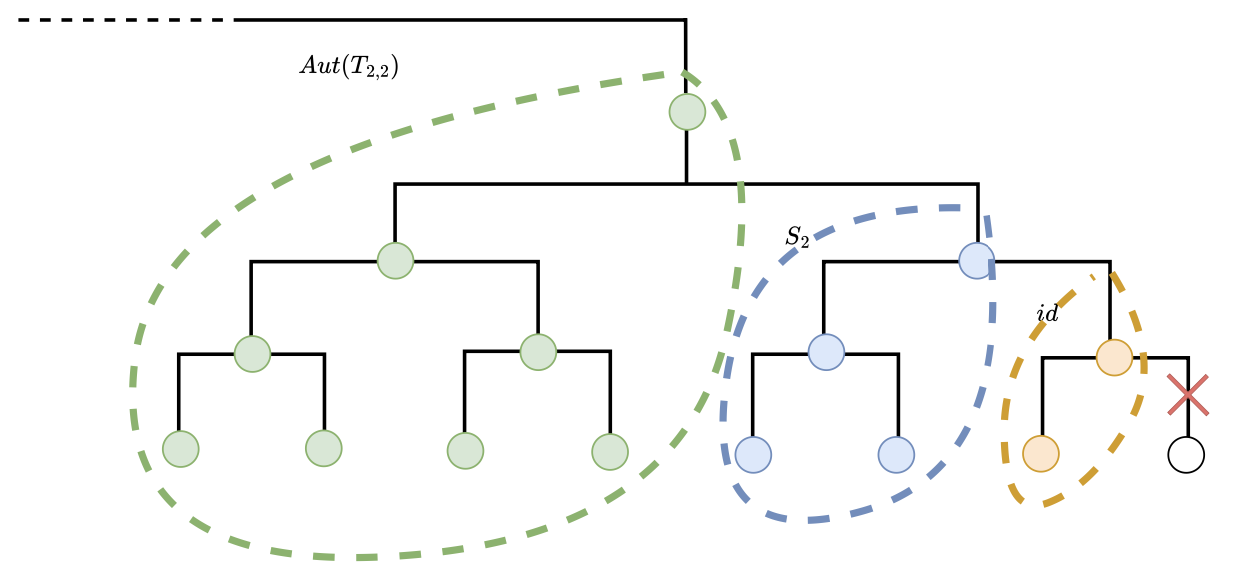}
\caption{Sketch of the proof of Proposition \ref{Prop. 9} with $r=h$.}
\label{Fig. 4a}
\end{figure}
\begin{figure}[!ht]
\centering
\includegraphics[width=0.75\columnwidth]{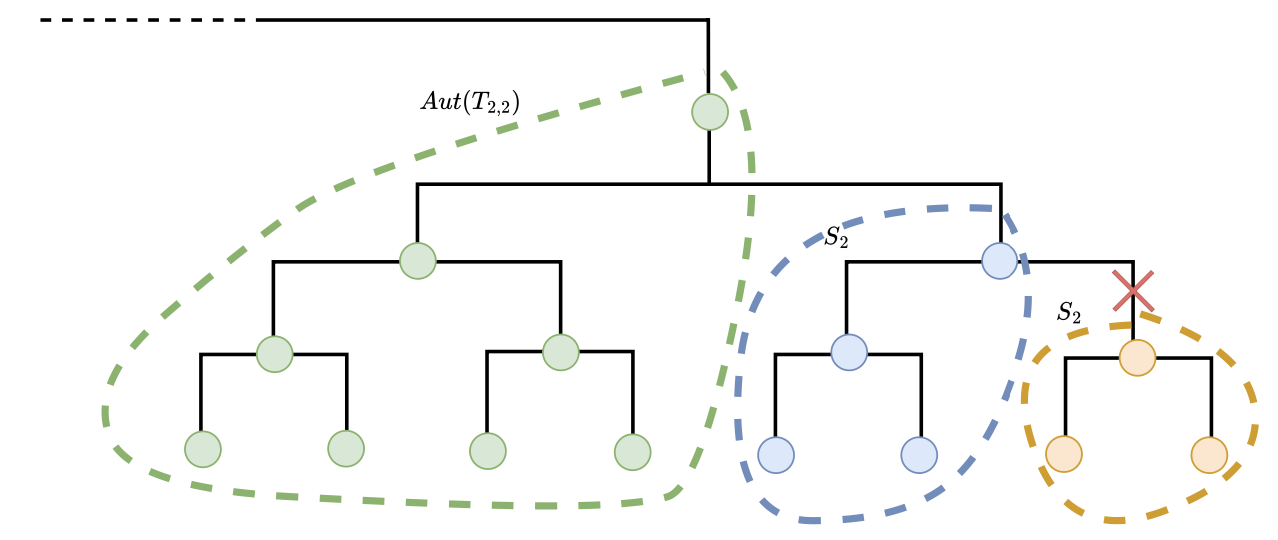}
\caption{Sketch of the proof of Proposition \ref{Prop. 9} with $r=h-1$.}
\label{Fig. 4b}
\end{figure}

\begin{figure}[!ht]
\centering
\includegraphics[width=0.8\columnwidth]{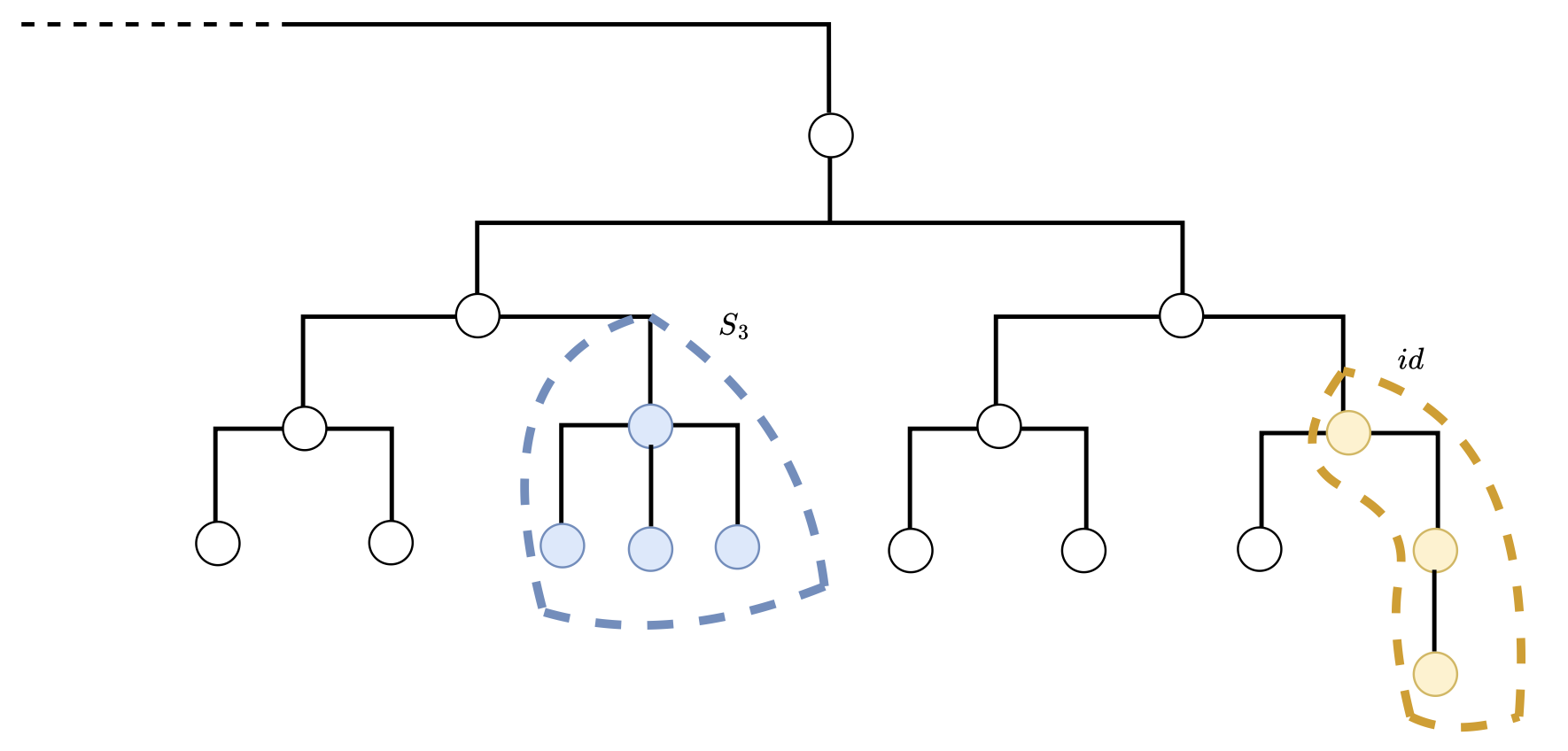}
\caption{Sketch of the proof of Proposition \ref{Prop. 10}.}
\label{Fig. Prop. 10}
\end{figure}

\noindent\textit{Proof.} Without loss of generality we can suppose that the deleted edge is in the rightmost branch of the tree. If $r=h$ then the rightmost bottom level is given by a path of length two plus a shadow node and a tree $T'_{2,1}$ given by $T_{2,1}$ with an extra edge at its root (see Fig.~\ref{Fig. 4a}) yielding a factor $\Aut(T'_{2,1})=\{id\}\times S_2\cong \Aut(T_{2,1})$. At level $h-1$ we have a factor $\Aut(T_{2,2})$ since the corresponding component is a full binary tree with $7$ nodes and an extra edge at its root, and we cannot swap this sub-branch of $\Gamma'$ with the other sub-branch containing the deleted edge. Going up level by level we get components of the form $T'_{2,i}$, given, as before, by $T_{2,i}$ plus a pendent edge at its root, yielding factors $\Aut(T_{2,i})$ for $i=3,\dots, h-1$. If $r=h-1$ we can reason analogously, noting that in the rightmost branch there is a disconnected component with symmetry group $\Aut(T_{2,1})\cong S_2$ (see Fig.~\ref{Fig. 4b}) which gives the first factor in the second point, then there is the associated component given by $T'_{2,1}$ as before, and from level $h-2$ we have the same subgraphs as in the first point. If $r<h-1$ the first factor in $\Aut(\Gamma')$ is a $\Aut(T_{2,r})$ since the deletion of an edge at level $r$ generates a connected component in $\Gamma'$ equal to $T_{2,r}$. The remaining factors are obtained as in the first point, starting from the component $T'_{2,r}$ given by $T_{2,r}$ plus an extra edge at its root, and going up level by level. \hfill $\Box$
\begin{Proposition}\label{Prop. 10} A pendent edge added to $T_{2,h}$ at level $0<r\leq h$ generates a graph $\Gamma'$ such that:
\begin{enumerate}
    \item If $r=h$ then $\Aut(\Gamma')\cong \Aut(T_{2,1})\times \Aut(T_{2,2})\times\dots \times \Aut(T_{2,h-1})$.
    \item If $r=h-1$ then $\Aut(\Gamma')\cong S_3\times \Aut(T_{2,1})\times \Aut(T_{2,2})\times \dots \times \Aut(T_{2,h-1})$.
    \item If $r<h-1$ then $\Aut(\Gamma')\cong \Aut(T_{2,r-1})\times \Aut(T_{2,r-1})\times \Aut(T_{2,r})\times \dots \times \Aut(T_{2,h-1})$.
\end{enumerate}
\end{Proposition}
\noindent\textit{Proof.} The first and last points are the analogue of the first and second points in Proposition \ref{Prop. 9}. The second point is obtained noting that at level $h-1$ there are three nodes of degree one in $\Gamma'$ that can be permuted by any element of $S_3$ (see Fig.~\ref{Fig. Prop. 10}). \hfill $\Box$
\begin{Proposition}\label{Prop. 11}
Removing an edge from $K_n$ generates a graph $\Gamma'$ with 
\begin{equation}
\Aut(\Gamma')\cong S_2\times \Aut(K_{n-2})=S_2\times S_{n-2}\,.
\end{equation}
\end{Proposition}
\noindent\textit{Proof.} Removing an edge from $K_n$ yields two nodes $v,w$ in $\Gamma'$ of degree $n-2$ and any automorphism must fix them and acts on the remaining nodes as a permutation of $S_{n-2}$. \hfill $\Box$ 
\begin{Proposition}\label{Prop. 12}
Adding a pendent edge to $K_n$ generates a graph $\Gamma'$ with $\Aut(\Gamma')\cong \Aut(K_{n-1})$.
\end{Proposition}
\noindent\textit{Proof.} For every $\alpha\in \Aut(\Gamma')$, the node of degree one and the node of degree $n+1$ in the added pendent edge must be fixed by $\alpha$ and $\alpha$ acts as permutation of $S_{n-1}$ on the remaining nodes.\hfill $\Box$

\subsection{QAOA circuits associated to the selected graphs and their perturbations}
We have seen that for graphs of type $K_n$ and $T_{2,h}$ we can compute closed formulas for the spectrum and the symmetry group, as well as closed formulas for their perturbed spectrum and symmetry group under the addition of shadow nodes or pendent edges, and the deletion of an edge. We have also remarked that for graphs of type $E_{q,n}$ and $R_{3,n}$ we cannot deduce such formulas due to the randomness intrinsic in their construction. In the next section we will generate specific graphs and compute directly the spectrum and the symmetry group, here we will discuss the effect of the perturbations introduced above on a QAOA circuit. 
\begin{Proposition}\label{Prop. 13} For a QAOA circuit with $p=1$ layers, we have that:
\begin{enumerate}
\item If $\Gamma'=(V',E)$ is obtained from $\Gamma$ by addition of $s\geq 1$ shadow nodes, then the mixer operator of the QAOA circuit for $\Gamma'$ is obtained from the mixer operator $U(\beta)$ of the QAOA circuit for $\Gamma$ as $U(\beta)\prod_{j=1}^se^{-i\beta\bm{\sigma}_x^{(j)}}$ and $U_{g_{\Gamma'}}=U_{g_\Gamma}$.
\item If $\Gamma'=(V,E')$ is obtained from $\Gamma$ by deletion of an edge, then the mixer operator of the QAOA circuit for $\Gamma'$ is $U(\beta)$ as the mixer operator of the QAOA circuit associated to $\Gamma$, and $g_{\Gamma'}(\bm{\sigma})=\sum_{(u,v)\in E'}(\mathbf{1}_4-\bm{\sigma}_z^{(u)}\otimes \bm{\sigma}_z^{(v)})$, thus $U_{g_{\Gamma'}(\bm{\sigma})}(\gamma)=e^{-i\gamma g_{\Gamma'}(\bm{\sigma})}$.
\item If $\Gamma'=(V',E')$ is obtained from $\Gamma$ adding a pendent edge $(u,v')$ with $u\in V$ and $v'\not \in V$, then the mixer operator of the QAOA circuit for $\Gamma'$ is obtained from the mixer operator $U(\beta)$ of the QAOA circuit for $\Gamma$ as $U(\beta)e^{-i\beta\bm{\sigma}_x^{(v')}}$ and $U_{g_{\Gamma'}}(\gamma)=U^\Gamma(\gamma)e^{-i\gamma(1-\bm{\sigma}_z^{(u)}\otimes \bm{\sigma}_z^{(v')})}$.
\end{enumerate}
\end{Proposition}
\noindent\textit{Proof.} A straightforward calculation in the mixer operator associated to $\Gamma$ plus $s$ shadow nodes yields
\begin{equation}
e^{-i\beta\sum_{j=1}^{n+s}\bm{\sigma}_x^{(j)}}=e^{-i\beta\sum_{j=1}^n\bm{\sigma}_x^{(j)}}e^{-i\beta\sum_{j=n}^{n+s}\bm{\sigma}_x^{(j)}}= U(\beta)\prod_{j=1}^se^{-i\beta\bm{\sigma}_x^{(j)}}\,.
\end{equation}
The first point is then proved observing that $\Gamma$ and $\Gamma'$ have the same set of edges. The second point is obtained also with a similar straightforward calculation and by definition of cost operator. The last point follows from the fact that the mixer operator in the QAOA circuit associated to $\Gamma'$ has an extra term $e^{-i\beta\bm{\sigma}_x^{(v')}}$ corresponding the degree-one node in the pendant edge, and its cost operator has an extra term $e^{-i\gamma(\mathbf{1}_4-\bm{\sigma}_z^{(u)}\otimes\bm{\sigma}_z^{(v')} )}$ corresponding to the added pendent edge. \hfill $\Box$
\\

\begin{figure}[!ht]
\centering
\includegraphics[width=0.9\columnwidth]{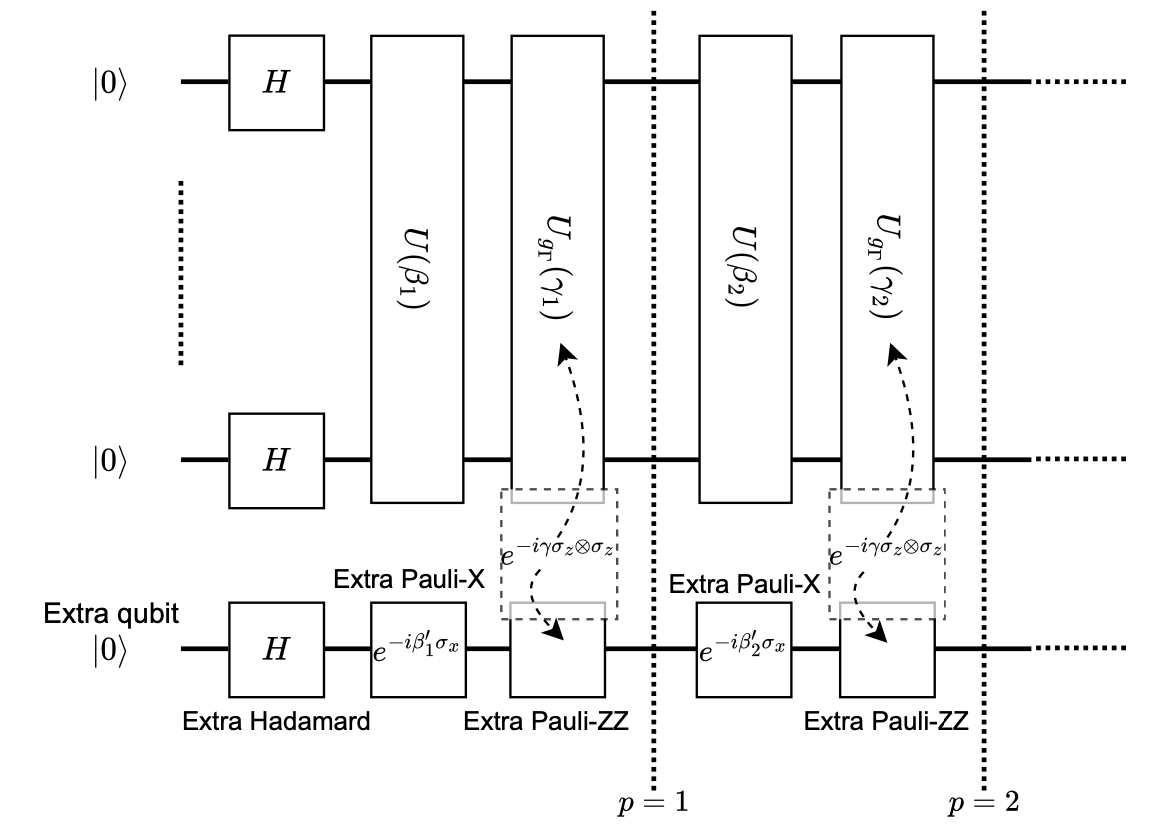}
\caption{QAOA circuit with $p=2$ for a generic perturbation of a graph $\Gamma=(V,E)\to \Gamma'=(V\cup \{\overline{v}\}, E\cup \{\overline{e}\})$. There is an extra $H$ gate for the qubit corresponding to $\overline{v}$ at $p=1$, then, for each layer, there is an extra Pauli-$x$ rotation corresponding to the action of the mixer operator on $\overline{v}$, then the Pauli-$zz$ gate acting on the edge $\overline{e}$ connecting $u\in V$ to $\overline{v}$.}
\label{Fig. 12}
\end{figure}

We can immediately generalize Proposition \ref{Prop. 13} for $p>1$: for every layer the operators corresponding to the perturbed graphs must be changed in the same way as they where changed for $p=1$. Concretely, once we have the QAOA circuit with $p\geq 2$ layers associated to a graph $\Gamma$, we can obtain the QAOA circuit, with the same $p$, corresponding to one of its perturbations $\Gamma'$ either by adding a qubit and a rotation $e^{-i\beta\bm{\sigma}_x}$, or by keeping the same number of qubits and remove a rotation $e^{-i\gamma\bm{\sigma}_z\otimes\bm{\sigma}_z}$ from the circuit, or by adding a qubit together with a rotation $e^{-i\beta\bm{\sigma}_x}$ and a rotation $e^{-i\gamma\bm{\sigma}_z\otimes\bm{\sigma}_z}$, as illustrated in Fig.~\ref{Fig. 12}.
%

%
%
\section{Implementation and Experimental Setup}
\label{sect:implementation}
As discussed in the previous sections, we will solve the MaxCut problem using the QAOA on some graphs of interest and their perturbations. Here we present the dataset used in the experiments, and the experimental methodology.

\subsection{Dataset}
\begin{figure}[!ht]
\centering
\includegraphics[width=1\columnwidth]{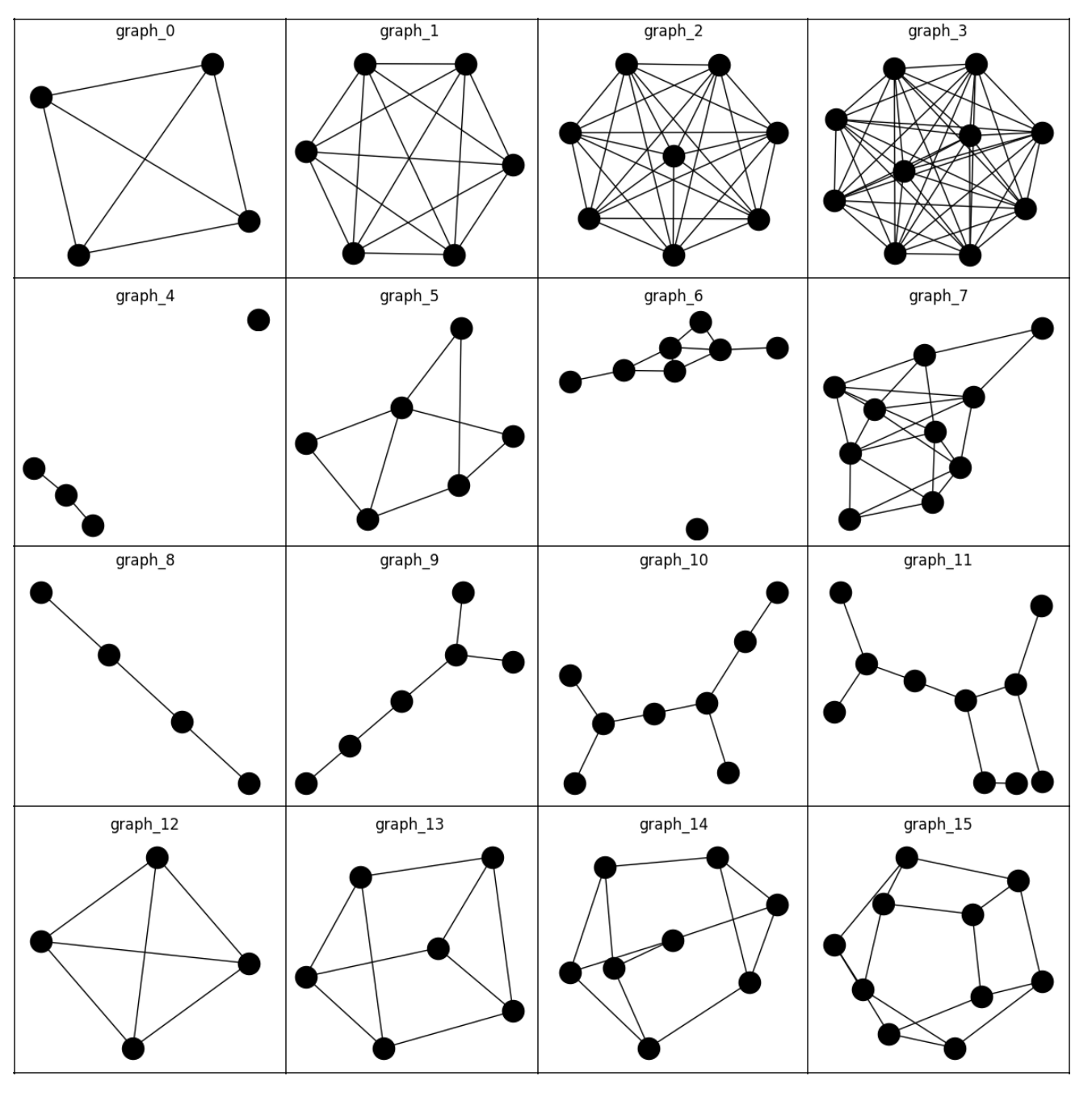}
\caption{Selection of graphs in the dataset chosen. In particular \texttt{graph\_i} for $i=0,1,2,3$ are complete graphs, \texttt{graph\_i} for $i=4,5,6,7$ are Erd\H{o}s-R\'enyi graphs, \texttt{graph\_i} for $i=8,9,10,11$ are full binary trees, and \texttt{graph\_i} for $i=12,13,14,15$ are random 3-regular graphs.}
\label{Fig. graph_select}
\end{figure}

We have considered $K_n$ with $n\in \{4,6,8,10\}$, $T_{2,h}$ with $h\in \{1,2\}$, $E_{q,n}$ with $q=0.5$ and $n\in \{4,6,8,10\}$, $R_{3,n}$ with $n\in \{4,6,8,10\}$. To strengthen our experimental methodology, we have also considered full $r$-rary trees with $r=2$ and $n\in \{4,6,8,10\}$ nodes, which we will denote as $\tilde{T}_{2,n}$, and are binary trees with all non-leaf nodes having exactly two successors and all levels are full except for some rightmost position at height $h$, as shown in Fig.~\ref{Fig. graph_select}.
\begin{Remark}\label{Rem. 3}\normalfont The symmetries and spectrum of $\tilde{T}_{2,h}$ can be easily deduced by combining Proposition \ref{Prop. 9} and Proposition \ref{Prop. 10}. In particular such graphs are of the form $T_{2,h+1}-v$ with $v$ the bottom rightmost node, or of the form $T'_{2,h}$ where $T'_{2,h}$ has a pendent edge at its bottom rightmost node. In this way we can consider tree structures with an even number of nodes comparable with the other problem instances considered.
\end{Remark}
We generated the dataset containing the graphs described above within the \texttt{Python v.3.11} environment using the library \texttt{NetworkX v.3.3}. The corresponding data-generation code together with saved problem instances is available in the repository associated to this paper, which can be accessed at \url{https://github.com/NesyaLab/Papers_with_code}.

\subsection{Experimental setup and evaluation criteria}
For each graph in the dataset we have considered their four perturbations: addition of $s\in \{1,2\}$ shadow nodes, addition of a pendent edge, and deletion of an edge. For each of these graphs we executed the QAOA on a simulator using \texttt{Qiskit v.0.45.2} on a \texttt{qiskit-aer v.0.13.2} backend. For robustness and reproducibility we considered three different seeded executions, from which we extracted the performance metrics taking averages and standard deviations. When closed formulas about the automorphisms of the graphs were unavailable we used the module \texttt{pynauty v.2.8.6} to compute them. For spectral calculations we used the module \texttt{numpy v.1.26.3}. All the requirements and associated code are available in the companion repository of the paper \url{https://github.com/NesyaLab/Papers_with_code}.

Since the results about graphs $T_{2,h}$ and $\tilde{T}_{2,h}$ are practically the same, we experimentally focused only on $\tilde{T}_{2,h}$ in order to keep a coherent number of nodes among all the simulations. The corresponding performances were measured by solving the MaxCut problem on the selected graphs and their perturbations. The metrics are summarized below.
\begin{itemize}
\item The mean approximation ratio 
\begin{equation}
\mu_{p,\Gamma}:=\mathbb{E}(AR(p,\Gamma))\,,
\end{equation}
where the mean is taken across each graph type with respect to either the number of nodes $n$ or the number of layers $p$.
\item The quotient 
\begin{equation}
I':=\frac{\mu_{p,\Gamma}}{\mu_{p,\Gamma'}}\,,
\end{equation}
where $\Gamma'$ is a perturbation of the problem instance $\Gamma$.
\item The symmetry index 
\begin{equation}
I_{\Sym}:=\frac{g_\Gamma(z^*)|\Aut(\Gamma')|}{g_{\Gamma'}(z^*)|\Aut(\Gamma)|}\,,
\end{equation}
where $g_{\Gamma}(z^*)$, $g_{\Gamma'}(z^*)$ is the optimal solution of the MaxCut problem with graph $\Gamma$, $\Gamma'$, respectively.
\item The approximate index 
\begin{equation}
I'_{\Sym}:=\frac{\mu_{p,\Gamma}|\Aut(\Gamma')|}{\mu_{p,\Gamma'}|\Aut(\Gamma)|}\,.
\end{equation}
%
%
\end{itemize}
%

%
%
\section{Numerical Results}
\label{sect:results}
In this section we present our main findings: we calculated the spectrum and symmetry group for each of the considered graphs, we calculated the MaxCut value, the relationship between the graph perturbations and the metrics considered, as well as the experimental performance of the QAOA with respect to these metrics. From these theoretical results and experiments we deduced some interesting heuristics that can be applied to solve high-dimensional problems with a QAOA approach, by leveraging symmetry and spectral information.

If $\varrho(K_n)$ is the spectral radius of the generated graphs $K_n$, if $\varrho(K'_n)$ is the spectral radius of the graph $K'_n$ obtained from $K_n$ adding $s>0$ shadow nodes, then $\varrho(K_n)=\varrho(K'_n)=n-1$ by Corollary \ref{Cor. 2} and Proposition \ref{Prop. 1}. With a similar argument we can deduce also that $\varrho(\Gamma)=\varrho(\Gamma')$ when $\Gamma'$ is obtained from $\Gamma\in \{\tilde{T}_{2,h},R_{3,n}\}$ by adding $s>0$ shadow nodes, indeed the addition of shadow nodes yields eigenvalues in $\mathbf{a}(\Gamma')$ smaller than $\varrho(\Gamma)$. For what concerns $E_{q,n}$ we note that $E_{0.5,4}$ is $P_3$ plus a shadow node (Fig.~\ref{Fig. graph_select} second row first column), so, by direct calculation, $\varrho(P_3)=\sqrt{2}$, and the addition of another shadow doesn't change the spectral radius' value. Moreover $E_{0.5,6}$ and $E_{0.5,10}$ are connected (Fig.~\ref{Fig. graph_select} second row second and last columns) and Proposition \ref{Prop. 1} also applies in this case. In conclusion $E_{0.5,8}$ decomposes as a connected graph on seven nodes and a shadow node (Fig.~\ref{Fig. graph_select} second row third column), so we can reason as in Proposition \ref{Prop. 12}, thus, for all the graphs $\Gamma$ considered, it holds the following result.
\begin{Proposition}\label{Prop. 14} For each generated graph $\Gamma$ and each associated perturbation $\Gamma'$ it holds $\varrho(\Gamma)=\varrho(\Gamma')$.
\end{Proposition} 

Using the results in Sect.~\ref{sect:QAOA_circuit} in combination with automoprhism-finding routines, we obtain in Table~\ref{Tab. 1} the size of $\Aut(\Gamma)$ and $\Aut(\Gamma')$ for the graphs considered and their perturbations $\Gamma'\in \{\Gamma^1,\Gamma^2,\Gamma^{del},\Gamma^{pen}\}$, where $\Gamma^s$ for $s\in \{1,2\}$ corresponds to the perturbation with $s$ shadow nodes, $\Gamma^{pen}$ corresponds to the addition of a pendent edge to $\Gamma$, and $\Gamma^{del}$ corresponds to the perturbation of $\Gamma$ given by the deletion of an edge.
\begin{table*}[!t]
    \centering
	   \captionof{table}{Number of automorphisms of the considered graphs and their perturbations. An ordered set $(a,b,c,d)$ corresponds to a graph with $n$ nodes with $n$ orderly taken from $(4,6,8,10)$. The maximal element is taken across the graphs $\Gamma'$ obtained from $\Gamma$ by deletion of an edge.}
    \normalsize
    \begin{adjustbox}{width=\textwidth,center}
    \begin{tabular}{lccccc}
        \toprule
         & $K_n$ & $E_{0.5,n}$ & $\tilde{T}_{2,h}$ & $R_{3,n}$  \\
        \midrule
        $|\Aut(\Gamma)|$ & $(4!, 6!, 8!, 10!)$ & $(2, 2, 1, 1)$ & $(2, 2, 2, 4)$ & $(24, 12, 12, 20)$  \\
        $|\Aut(\Gamma^1)|$ & $(4!, 6!, 8!, 10!)$ & $(4, 2, 2, 1)$ &  $(2, 2, 2, 4)$ &  $(24, 12, 12, 20)$\\
        $|\Aut(\Gamma^2)|$ & $(2\times 4!, 2\times 6!, 2\times 8!, 2\times 10!)$ & $(12, 4, 6, 2)$ & $(4, 4, 4, 8)$ &  $(48, 24, 24, 40)$\\
        $|\Aut(\Gamma^{pen})|$ & $(4!/4, 6!/6, 8!/8, 10!/10)$ & $(4, 1, 1, 1)$ &  $(2, 6, 2, 12)$ &  $(6, 2, 4, 2)$ \\
        $\max|\Aut(\Gamma^{del})|$ & $(2\times 4!/12, 2\times 6!/30, 2\times8!/56, 2\times 10!/90)$ & $(4,6,4,2)$ &  $(8,12,12,16)$ &  $(4,4,4,4)$ \\
        \bottomrule
    \end{tabular}
    \label{Tab. 1}
    \end{adjustbox}
\end{table*}

We also report in Table~\ref{Tab. 2} the exact MaxCut values $g_\Gamma(z^*)$, $g_{\Gamma'}(z^*)$, computed with a brute-force approach for each graph and their perturbations, as well as the corresponding highest approximation ratio. 
\begin{table*}[!t]
    \centering
        \caption{MaxCut values and highest approximation ratio. An ordered set $(a,b,c,d)$ corresponds to a graph with $n$ nodes with $n$ orderly taken from $(4,6,8,10)$. The maximal elements are taken across the different number of layers $p$ considered.}
    \normalsize
    \begin{adjustbox}{width=\textwidth,center}
    \begin{tabular}{lccccc}
        \toprule
         & $K_n$ & $E_{0.5,n}$ & $\tilde{T}_{2,h}$ & $R_{3,n}$  \\
        \midrule
        $g(z^*)$ & $(4, 9, 16, 25)$ & $(2, 7, 7, 16)$ & $(3, 5, 7, 9)$ & $(4, 7, 10, 15)$  \\
        $\max\mu_{p,\Gamma}$ & $(0.998, 0.990, 0.959, 0.959)$ & $(0.997, 0.781, 0.831, 0.700)$ &  $(0.997, 0.781, 0.831, 0.700)$ &  $(0.999, 0.807, 0.843, 0.659)$\\
        $\max\mu_{p,\Gamma^1}$ & $(0.998, 0.980, 0.969, 0.952)$ & $(0.998, 0.849, 0.802, 0.784)$ & $(0.999, 0.848, 0.802, 0.784)$ &  $(0.997, 0.817, 0.818, 0.645)$\\
        $\max\mu_{p,\Gamma^2}$ & $(0.993, 0.969, 0.978, 0.948)$ & $(0.994, 0.855, 0.814, 0.658)$ &  $(0.994, 0.855, 0.814, 0.658)$ &  $(0.994, 0.855, 0.814, 0.658)$ \\
        $\max\mu_{p,\Gamma^{pen}}$ & $(0.999, 0.986, 0.981, 0.972)$ & $(0.998, 0.815, 0.843, 0.643)$ &  $(0.998, 0.815, 0.843, 0.643)$ &  $(0.998, 0.815, 0.843, 0.643)$ \\
        $\max\mu_{p,\Gamma^{del}}$ & $(1.000, 0.999, 0.996, 0.994)$ & $(1.000, 0.869, 0.849, 0.667)$ &  $(1.000, 0.869, 0.849, 0.667)$ &  $(1.000, 0.869, 0.849, 0.667)$ \\
        \bottomrule
    \end{tabular}
    \label{Tab. 2}
    \end{adjustbox}
\end{table*}
\begin{Remark}\label{Rem. 4}\normalfont If $\Gamma'$ is obtained from $\Gamma$ by the addition of $s>0$ shadow nodes, clearly the exact MaxCut value $g_\Gamma(z^*)$ with $z^*\in \{0,1\}^n$ is the same as the exact MaxCut value $g_{\Gamma'}(z'^*)$ for the perturbed graph with $z'^*\in \{0,1\}^{n+s}$. On the other hand the addition of a pendent edge or the deletion of an edge change the corresponding MaxCut value. As before if $\Gamma^{pen}=(V',E')$ is obtained from $\Gamma=(V,E)$ by the addition of a pendent edge connecting $u\in V$ to $v\in V'\setminus V$, then either $g_{\Gamma^{pen}}(z^{*pen})=g_\Gamma(z^*)+1$ with $z^{*pen}\in \{0,1\}^{n+1}$ when $u\in S$ and $v\in V'\setminus S$ are nodes in the two distinct elements of the MaxCut partition $\{S,V'\setminus S\}$, or $g_{\Gamma^{pen}}(z^{*pen})=g_\Gamma(z^*)$ when $\{u,v\}\subset S$ or $\{u,v\}\subset V'\setminus S$. Likewise given the perturbation $\Gamma^{del}=(V,E')$ we have either $g_{\Gamma^{del}}(z^{*del})=g_\Gamma(z^*)-1$ when the deleted edge connected two nodes $u\in S$ and $v\in V\setminus S$, or $g_{\Gamma^{del}}(z^{*del})=g_\Gamma(z^*)$ when the edge is used to connect nodes in the same element of the MaxCut partition. Since the addition of a pendent edge, and the deletion of an edge change the corresponding MaxCut problem we have reasoned as follows: if $\beta_1,\dots, \beta_n$ and $\gamma_1,\dots, \gamma_m$ are initial angles where $\beta_i=\beta$ for each $i=1,\dots, n$ and $\gamma_j=\gamma$ for each $j=1,\dots, m$, and if $\beta^*_1,\dots, \beta^*_n, \gamma_1^*,\dots, \gamma_m^*$ are the corresponding optimal parameters of the MaxCut problem on $\Gamma^{pen}$ achieved by a QAOA simulation with $p$ layers, then to compute $AR(p,\Gamma)$ we considered $\beta^*_1,\dots, \beta^*_{n-1},\gamma_1^*,\dots, \gamma^*_{m-1}$ only the parameters corresponding to the nodes and edges of $\Gamma$. Since for $\Gamma^{del}$ we have $g_{\Gamma^{del}}(z^{*del})\leq g_{\Gamma}(z^*)$ we kept the parameters corresponding to $AR(p,\Gamma^{del})$ by extending them arbitrarily, which we noticed doesn't affect $AR(p,\Gamma)$ significantly.
\end{Remark}
By evaluating the spectral radius and the number of symmetries for each graphs we can deduce the following heuristics.
\begin{Heuristic}\label{Heu. 1} We have from a classical result \cite{Tre12} that 
\begin{equation}
g_\Gamma(z^*) \leq \frac{1}{2} + \frac{1}{2 }\varrho(\Gamma)\,,
\end{equation}
which, for the graphs considered and their perturbations, can also be written as
\begin{equation}
g_\Gamma(z^*) \leq \frac{1}{2} + \frac{1}{2}\varrho(\Gamma')\,.
\end{equation}
Moreover the inequality $g_\Gamma(z^*) \leq g_\Gamma(z)$ allows to write, for $p \rightarrow \infty$, the asymptotics
\begin{equation}
F_{p,\Gamma}(\beta^*,\gamma^*) \sim g_\Gamma(z^*) \leq \frac{1}{2} + \frac{1}{2}\varrho(\Gamma')\,,
\end{equation}
which grants us an upper bound to the MaxCut solution obtained by the QAOA. This is particularly useful in real applications where the approximation ratio $AR(p,\Gamma)$ cannot be evaluated being the MaxCut value $g(z^*)$ unknown. It's important to highlight that especially for large graphs $\varrho(\Gamma')$ might be easier to compute compared to $\varrho(\Gamma)$.
\end{Heuristic}

\begin{figure}[!ht]
\centering
\includegraphics[width=\columnwidth]{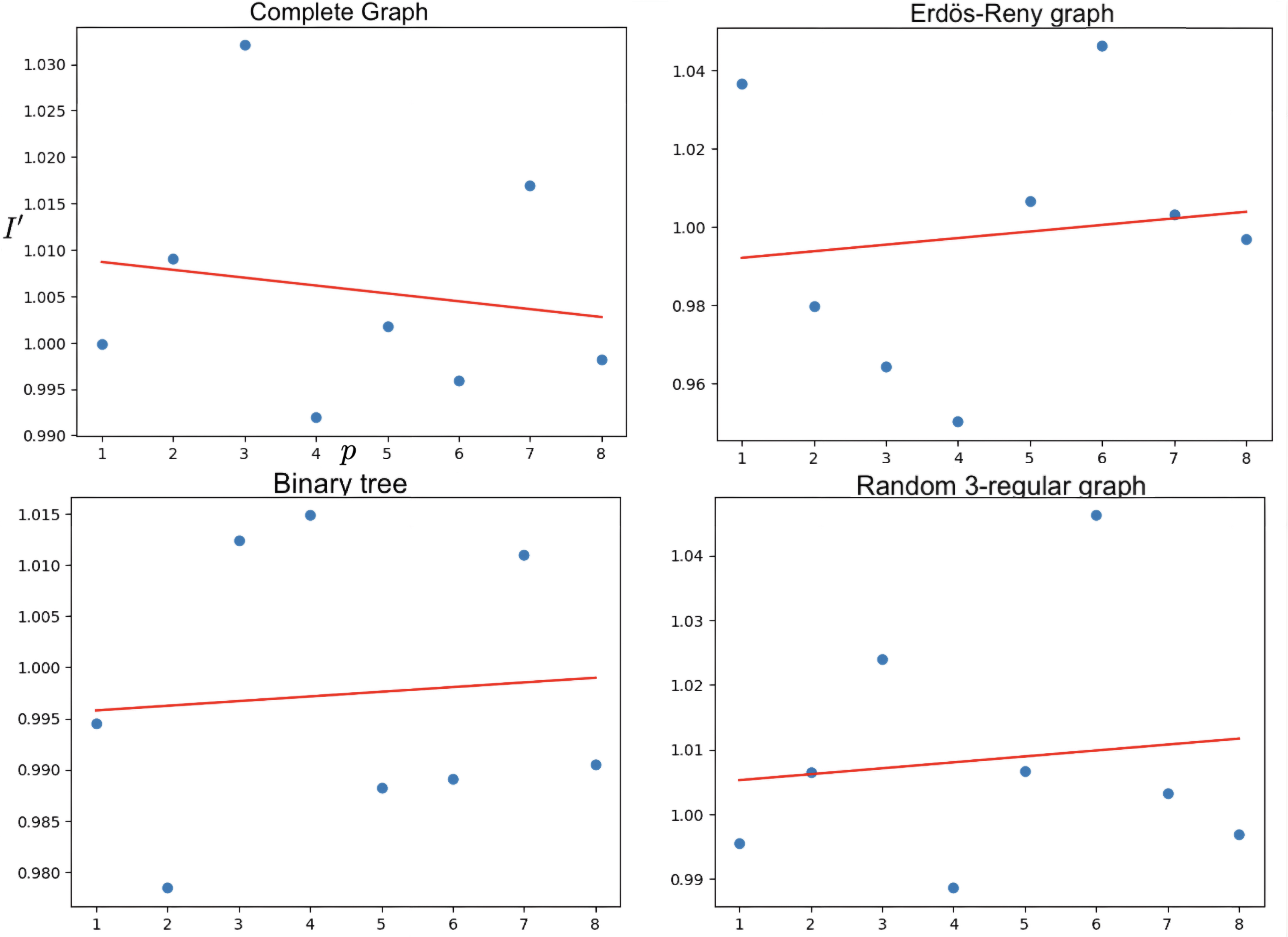}
\caption{Variation of the quotient $I'$ on a selected graph per graph class across different $p$ values.}
\label{Fig. 8}
\end{figure}

\begin{figure}[!ht]
\centering
\includegraphics[width=\columnwidth]{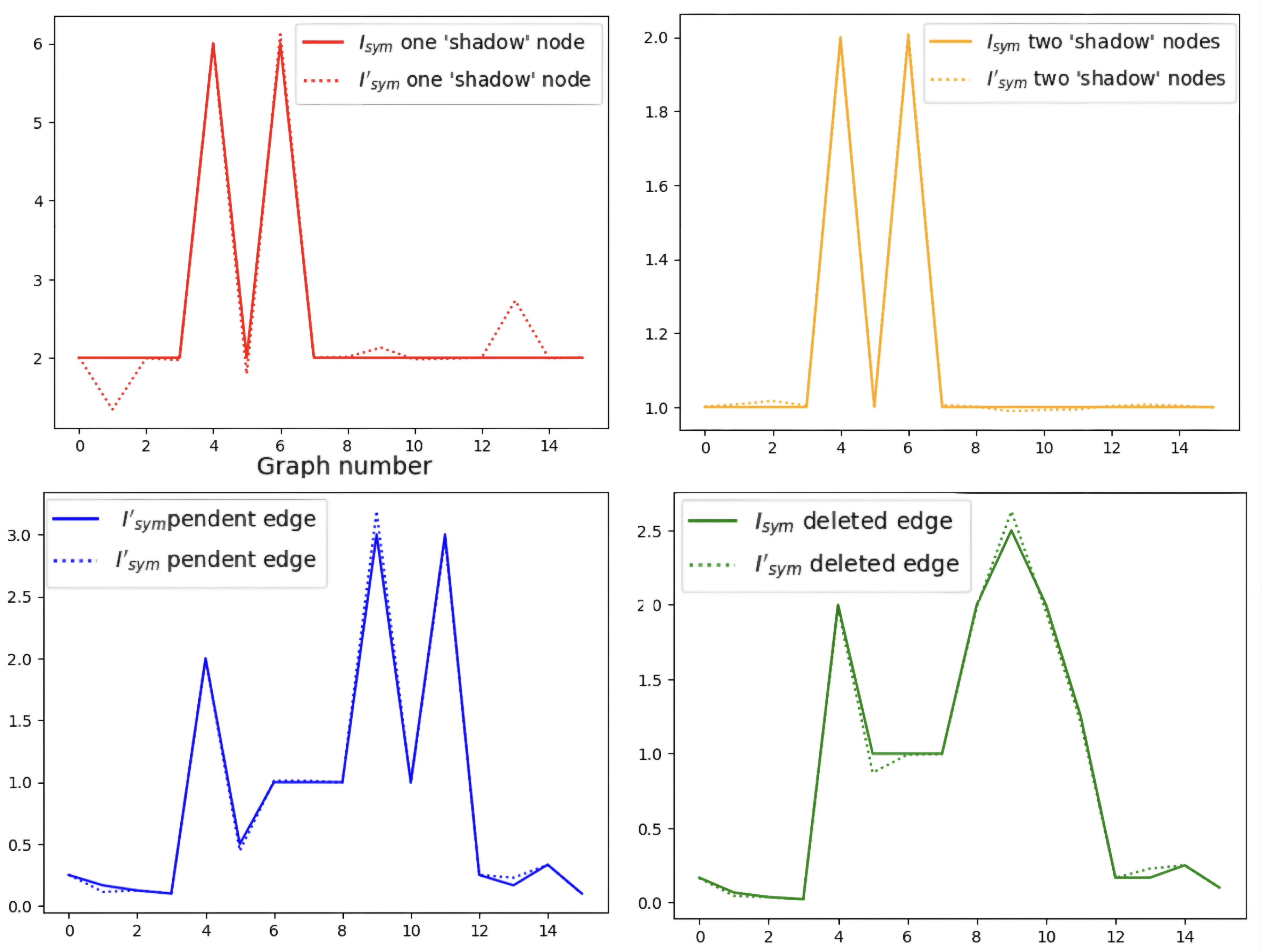}
\caption{Exact symmetry index $I_{sym}$ and approximate symmetry index $I'_{sym}$ for each graph type and their perturbations, where \texttt{graph\_i} for $i=0,1,2,3$ are complete graphs, \texttt{graph\_i} for $i=4,5,6,7$ are Erd\H{o}s-R\'enyi graphs, \texttt{graph\_i} for $i=8,9,10,11$ are full binary trees, and \texttt{graph\_i} for $i=12,13,14,15$ are random 3-regular graphs.}
\label{Fig. 7}
\end{figure}

\begin{figure}[!ht]
\centering
\includegraphics[width=\columnwidth]{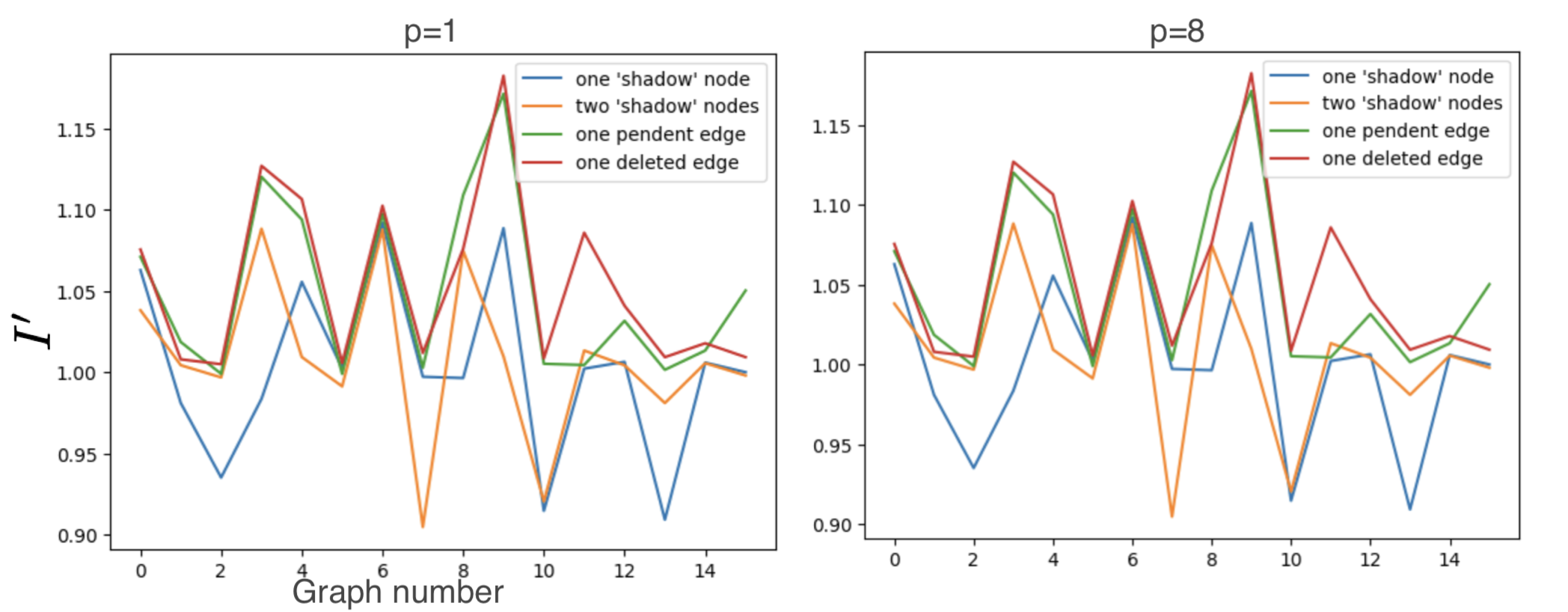}
\caption{Quotient $I'$ for each graph type and their perturbations, where \texttt{graph\_i} for $i=0,1,2,3$ are complete graphs, \texttt{graph\_i} for $i=4,5,6,7$ are Erd\H{o}s-R\'enyi graphs, \texttt{graph\_i} for $i=8,9,10,11$ are full binary trees, and \texttt{graph\_i} for $i=12,13,14,15$ are random 3-regular graphs.}
\label{Fig. 9}
\end{figure}

Having upper bounds $M$ for the MaxCut is often important, not only to get a substitute metric of the form $F_p(\beta^*,\gamma^*)/M$ when $g(z^*)$ is unavailable, but also to get a certificate for the solution. It is often easy for a graph $\Gamma$ and a cut value of $c$ to certify that $\max g_\Gamma\geq c$ since it suffices to exhibit a single bipartition $z$ of $\Gamma$ such that $g_\Gamma(z)>c$. In contrast it is unclear how to certify statements of the form $\max g_\Gamma \leq c$ since to verify this case we must rule-out an exponential number of possible bipartitions. 
Many results in this sense (e.g. \cite{Veg07}) show that large classes of classical algorithm based on Linear Programming relaxations fail to distinguish such instances.
\begin{Heuristic}\label{Heu. 2} The two metrics $I'=I'(p,\Gamma,\Gamma')$, $ I'_{sym}=I'_{sym}(p,\Gamma,\Gamma')$  remain constant for the considered number of layers $p\in \{1,2,3,4,5,6,7,8\}$ (up to a negligible error, cf. Propositions \ref{Prop. 9}, \ref{Prop. 10}, \ref{Prop. 11}, \ref{Prop. 12} and Figs.~\ref{Fig. 8},~\ref{Fig. 7},~\ref{Fig. 9}).
\end{Heuristic}

We conjecture that the same heuristic holds for $p>8$, but we did not test this case.
\begin{Heuristic}\label{Heu. 3} The effect of adding a shadow node has no effect on the approximation ratio achieved by QAOA. The addition of a pendent edge and the deletion of an edge positively influences trees and random regular graphs.
\end{Heuristic}

\begin{figure}[!ht]
\centering
\includegraphics[width=0.7\columnwidth]{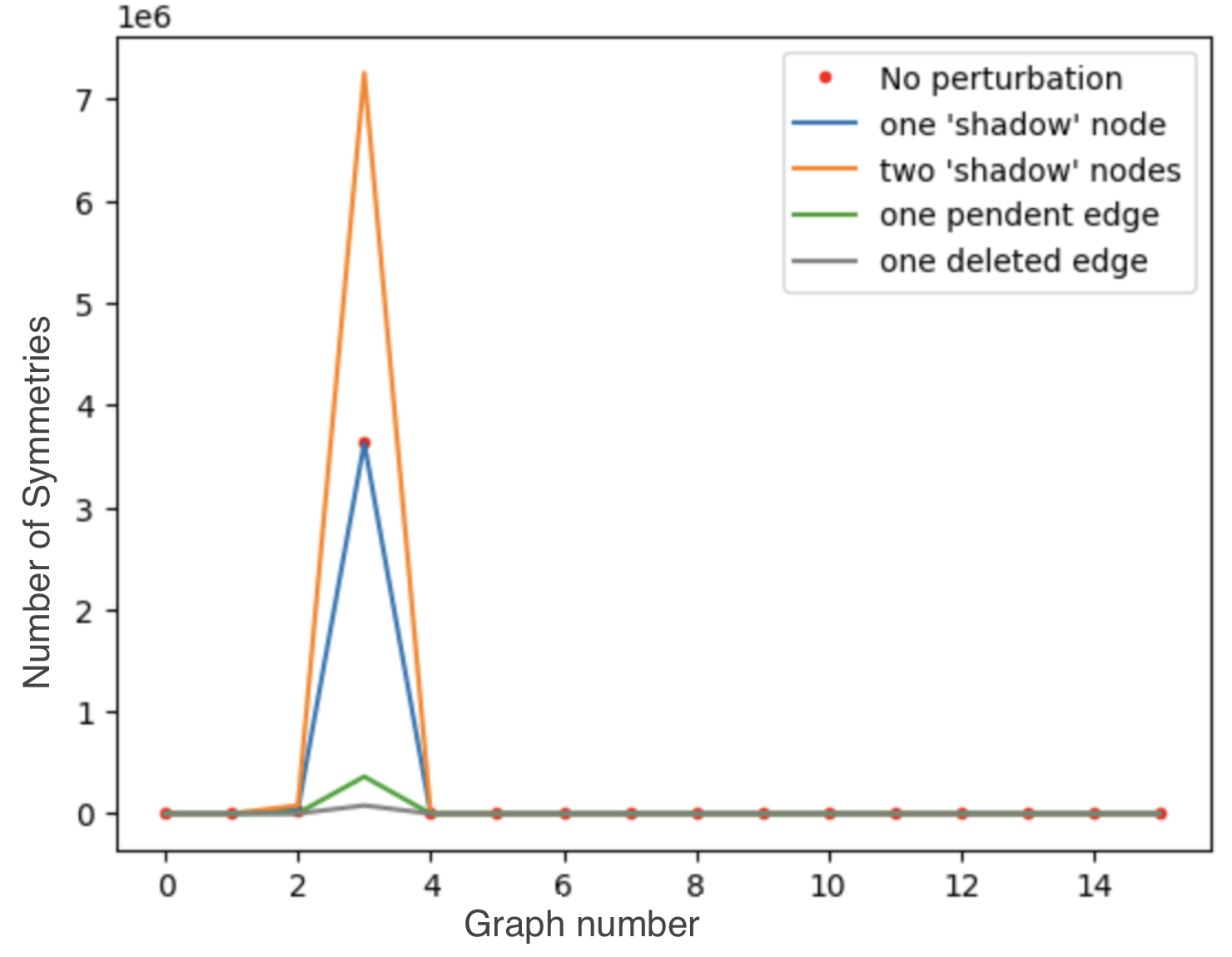}
\caption{Number of symmetries for each graph type and their perturbations, where \texttt{graph\_i} for $i=0,1,2,3$ are complete graphs, \texttt{graph\_i} for $i=4,5,6,7$ are Erd\H{o}s-R\'enyi graphs, \texttt{graph\_i} for $i=8,9,10,11$ are full binary trees, and \texttt{graph\_i} for $i=12,13,14,15$ are random 3-regular graphs.}
\label{Fig. 10}
\end{figure}

We conjecture that adding a single pendent edge yields the same effect of deleting a random edge in terms of approximation ratio achieved. In particular, if a graph has disconnected nodes they can be eliminated without QAOA performances decrease, but with a corresponding reduction in the number of qubits employed (Proposition \ref{Prop. 13}). Likewise, the deletion of few edges from a graph may not affect QAOA performances, yielding on the other hand a smaller circuits (with less \texttt{CNOT} gates in the implementation of the rotation $e^{-i\gamma\bm{\sigma}_z\otimes \bm{\sigma}_z}$). 
\begin{Heuristic}\label{Heu. 4} The quality of the symmetries in a graph matters: by adding two shadow nodes the number of symmetries is doubled (Proposition \ref{Prop. 7}), but the effect on the QAOA approximation ratio is negligible. Likewise the deletion or the addition of some edges could reduce the number of symmetries available (Propositions \ref{Prop. 9}, \ref{Prop. 10}, \ref{Prop. 11}, \ref{Prop. 12}), without hampering the QAOA performance (Fig.~\ref{Fig. 9} and Fig.~\ref{Fig. 10}).
\end{Heuristic}
%

%
%
\section{Conclusions}
\label{sect:conclusion}
Our study has provided valuable insights into the behavior of the QAOA when applied to graph optimization problems. Proposition \ref{Prop. 13} established a clear framework for understanding how perturbations in the graph structure translate into modifications in the QAOA circuit. We identified that certain alterations, such as the addition of shadow nodes or pendent edges, have predictable effects on the mixer operator and the associated cost function, thus offering a method to manipulate the QAOA's behavior in response to graph modifications.

Furthermore, the heuristics derived from our analysis offer practical guidance for leveraging these insights in real-world applications. We demonstrated that certain graph modifications, such as the addition of shadow nodes or the deletion of edges, have minimal impact on the QAOA's approximation ratio. This suggests avenues for optimizing QAOA performance by strategically adjusting the graph structure without compromising solution quality in terms of approximation ratio.

Moreover, our findings suggest that the computation of certain graph properties, such as the spectral radius, may be more tractable for modified graphs, thus facilitating the assessment of solution quality, otherwise inscrutable. Additionally, we observed that certain graph transformations, such as the addition or deletion of edges, can affect the symmetry properties of the graph, but these alterations do not significantly worsen the QAOA's performances.

In conclusion, our study advances the understanding of how graph perturbations influence the performance of the QAOA, offering practical strategies for optimizing algorithmic performance in graph optimization tasks. By leveraging these insights, researchers and practitioners can more effectively apply the QAOA to solve real-world optimization problems with greater efficiency and accuracy.

\section*{Acknowledgment}
The contribution of L. Lavagna, A. Ceschini and M. Panella in this work was in part supported by the ``NATIONAL CENTRE FOR HPC, BIG DATA AND QUANTUM COMPUTING'' (CN1, Spoke 10) within the Italian ``Piano Nazionale di Ripresa e Resilienza (PNRR)'', Mission 4 Component 2 Investment 1.4 funded by the European Union - {NextGenerationEU} - CN00000013 - CUP B83C22002940006.

\end{document}